# The Influence of Interior Noise on Just-Noticeable Speed Differences in Conventional and Electric Vehicles

**Zhenxian LI[1], Etienne Parizet[1], Claudio COLANGELI[2]**
1. INSA Lyon—LVA, France 2. Siemens Industry Software, Belgium

## Abstract

Electric vehicles (EVs) and internal-combustion-engine vehicles (ICEVs) differ fundamentally in their in-cabin acoustics, notably the attenuation or absence of engine-order content. Prior work reports associations between reduced engine sound, speed underestimation, and poorer speed maintenance; however, research on how EVs' new sound affects speed perception and control is scarce, and most newer studies focus on comfort and subjective pleasantness rather than speed perception. Addressing this gap, the present study uses a two-interval, two-alternative forced-choice (2AFC) paradigm to directly measure just-noticeable differences (JNDs) in speed under ICEV, EV, and silent conditions. Thirty participants performed a 2AFC task in which, on each trial, they viewed two first-person highway clips (reference vs. comparison) and indicated which appeared faster. Results from ANOVA and post-hoc tests indicate that at the 40 km/h reference speed participants showed no clear differences across sound conditions, whereas at 100 km/h there were marked differences in JND: mean values were 1.93 km/h (ICEV), 3.48 km/h (EV), and 5.15 km/h (silence). A psychoacoustic parameter analysis suggests that this effect is not explained by speed-dependent changes in loudness or sharpness; we interpret that RPM-related, clearly audible frequency shifts in ICEV provide the primary contributory cue. For EV NVH or artificial sound design, enhancing speed-contingent, trackable spectral cues while respecting comfort may help maintain drivers' ability to discriminate speed differences.

## Introduction

Accurate perception of ego-speed relies on the integration of multiple sensory cues, among which interior noise provides a crucial auditory reference. Classical work has demonstrated that attenuating vehicle noise systematically biases speed estimation: drivers tend to underestimate their speed and consequently drive faster when auditory cues are reduced or removed [1–3]. This effect has been repeatedly observed in controlled speed-matching and car-following tasks, where quieter cabins yield poorer speed maintenance and larger speed variability [4,5]. Such findings indicate that engine-related acoustic information is not merely a by-product of powertrain operation, but a functional feedback channel supporting perceptual calibration during driving [6].

The transition from internal combustion engine vehicles (ICEVs) to electric vehicles (EVs) fundamentally disrupts this auditory feedback channel. ICEVs provide a rich, speed- and load-dependent acoustic signature composed of engine orders, combustion harmonics, and broadband masking components that covary reliably with vehicle dynamics. These cues supply continuous information about acceleration, throttle input, and mechanical load, and are implicitly used by drivers to calibrate perceived speed and motion [7,8]. In contrast, EVs lack these characteristic harmonics and exhibit a markedly flatter and quieter interior acoustic profile, dominated instead by wind and tire–road interaction noise, and high-frequency tonal components from motor and inverter operation [9–12]. The absence of combustion-related structure not only removes a familiar mapping between sound and speed but also reduces natural masking, making previously insignificant noise components perceptually salient [12–14]. Studies on EV sound quality further show that this altered spectral balance weakens dynamic feedback, changes drivers' interpretation of vehicle state, and increases perceptual ambiguity regarding acceleration and speed [15–17]. Together, these findings suggest that EV acoustics may degrade the precision of internal motion estimation – raising the possibility that speed-change sensitivity intrinsically lower in EV interior soundscapes than in ICEVs.

Existing literature, however, has focused predominantly on mean speed bias – whether drivers go faster when sound is removed – while little is known about how reduced auditory cues influence the sensitivity to speed changes, i.e., the just-noticeable difference (JND) in speed. This gap is nontrivial: JND quantifies the minimal speed increment a driver can perceptually detect and thus constrains closed-loop speed control performance. If interior noise aids the discrimination of speed fluctuations, then EV quietness may directly degrade the perceptual resolution required for fine control, even when average speed appears unaffected [18].

Prior psychophysical studies provide partial insight. Visual-vestibular research shows that self-motion cues alter perceived speed without proportionally altering discrimination thresholds [19,20], while acceleration-perception work reports low JNDs only when strong dynamic cues are present [21]. Yet none of these studies manipulate interior auditory noise or examine how EV-like acoustic environments modify speed-change detectability under purely visual driving scenes. Moreover, existing speed-perception paradigms typically use steady-speed judgments or large speed differences, offering little resolution on whet her the quieter auditory profile of EVs fundamentally elevates the perceptual threshold for detecting small speed variations [22,23].

To address this gap, the present study quantifies drivers' JND for speed changes under three interior-sound conditions representative of contemporary vehicles: an "ICEV" condition, an "EV" condition, and a "Silence" condition. By systematically measuring JNDs, we directly test whether reduced auditory structure in EVs impairs the perceptual discrimination of longitudinal speed, beyond the well-documented bias effects on mean speed estimation. This approach provides a targeted psychophysical assessment of how interior noise contributes to the sensory foundation of speed control, and clarifies whether the acoustic quietness of EVs poses an intrinsic perceptual limitation for drivers.

## Methodology

### *Participants and Apparatus*

30 volunteer participants (students and staff from the university) took part in the experiment. All participants had normal or corrected-to-normal vision and no history of hearing problems. Each participant had at least 1 years of driving experience. Their exposure to vehicle types varied; most had predominantly ICEV driving experience, with some experience riding in EVs, while almost none had experience driving EVs.

The experiment was conducted using a fixed-base driving simulator. As shown in Figure 1, the setup included a large curved monitor (113 cm diagonal, 800R curvature), providing an approximately 113-degree field of view at a viewing distance of 60 cm. Participants sat at a standard driving position facing the screen, and the steering wheel and pedals were not used since only passive observation of speed was required. Audio stimuli were delivered through an RME Fireface UCX sound card and Sennheiser HD650 headphones. The audio playback and experiment control were run on a PC using MATLAB, and a VLC media player plugin was used to play the sound corresponding to each video condition. The sound pressure levels of two ears stimuli were calibrated based on real vehicle recordings and are summarized in Table 1.

Table 1. Sound pressure levels (dB(A)) delivered through audio system across experimental conditions. L and R denote the left and right ears, respectively.

| Condition | 40 km/h | | 100 km/h | |
|---|---|---|---|---|
| | (L) | (R) | (L) | (R) |
| ICEV | 53.5 | 53.1 | 63.7 | 63.6 |
| EV | 46.4 | 46.5 | 56.2 | 55.8 |

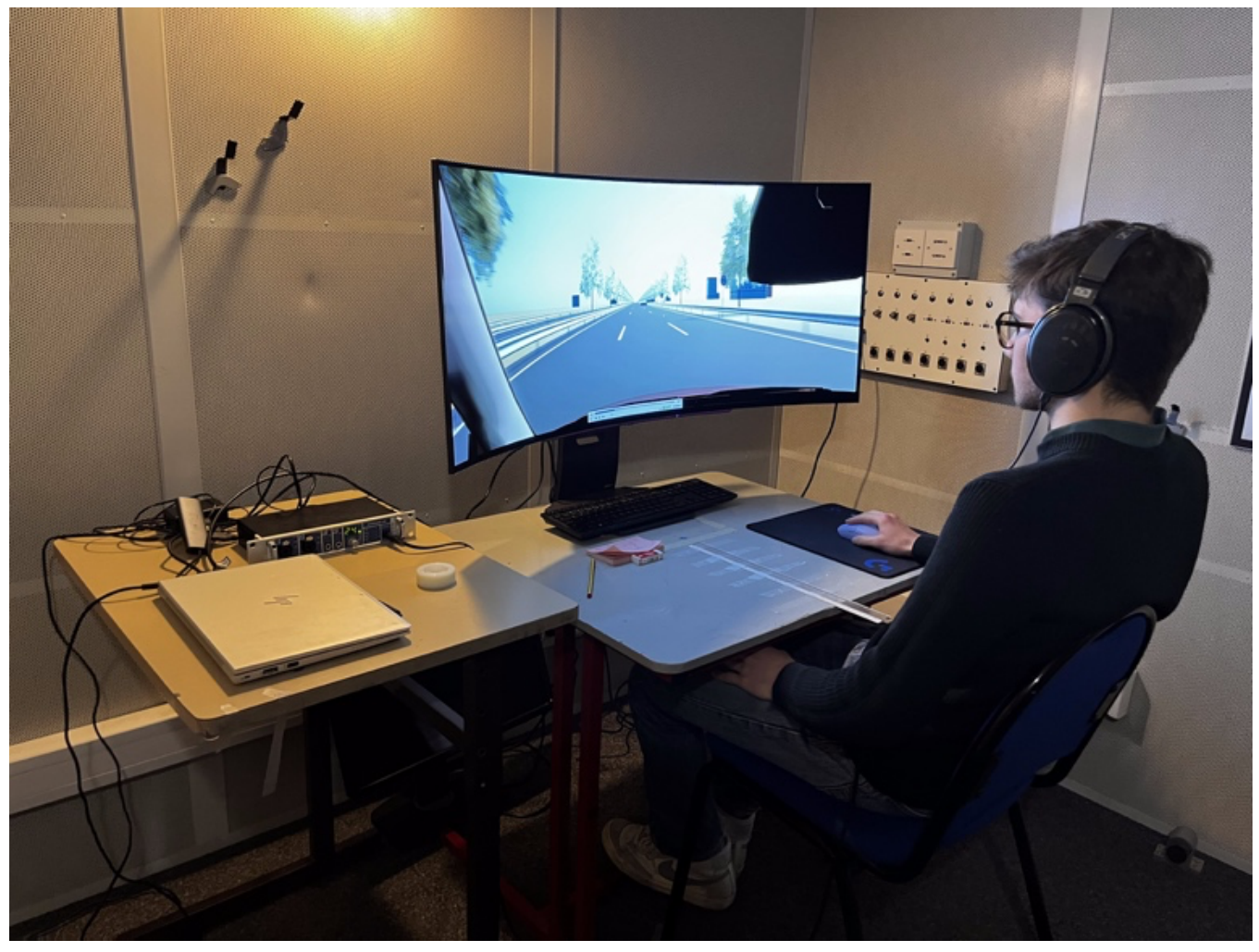

Figure 1. Scene of the driving simulator setup used in the experiment. This figure illustrates the participant's view of the simulation environment, including the large curved display and the driving scene.

### *Stimuli and Conditions*

The visual stimulus was a first-person perspective video of a highway driving scene generated using Siemens Prescan software. It depicted a French highway from the driver's viewpoint, with a straight road flanked by guardrails, roadside trees, and distant buildings (Figure 2). The video lasted approximately 7 seconds and was recorded at a constant speed for the reference stimulus. Each clip began from a different starting point along a straight highway segment featuring roadside trees, buildings, and standard highway signage. This design ensured that participants received sufficient visual information while preventing them from relying on memorized start–end scene cues, thereby avoiding additional biases in speed judgment. Two base speed levels were tested in separate blocks: a low speed of 40 km/h (city driving) and a high speed of 100 km/h (highway driving).

The interior sound stimuli were generated using the NVH sound simulation framework integrated into the driving simulator, based on Simcenter Testlab Real Time technology and co-simulated with the visual environment created in Simcenter Prescan [24]. The approach follows the Simcenter NVH Simulator framework， which combines measured vehicle NVH data with real-time auralization to reproduce realistic in-cabin sound signatures under controlled driving conditions [24–26]. Constant-speed scenarios were implemented using pre-calibrated NVH models representative of ICEV and EV powertrains, combined with identical tire and wind noise components, ensuring a coherent coupling between visual motion cues and auditory feedback.

The ICEV condition was generated using a NVH model of a Compact C-segment car equipped with Diesel engine. The EV condition was generated using representative NVH models derived from anonymized measurements of commercially available electric vehicles.

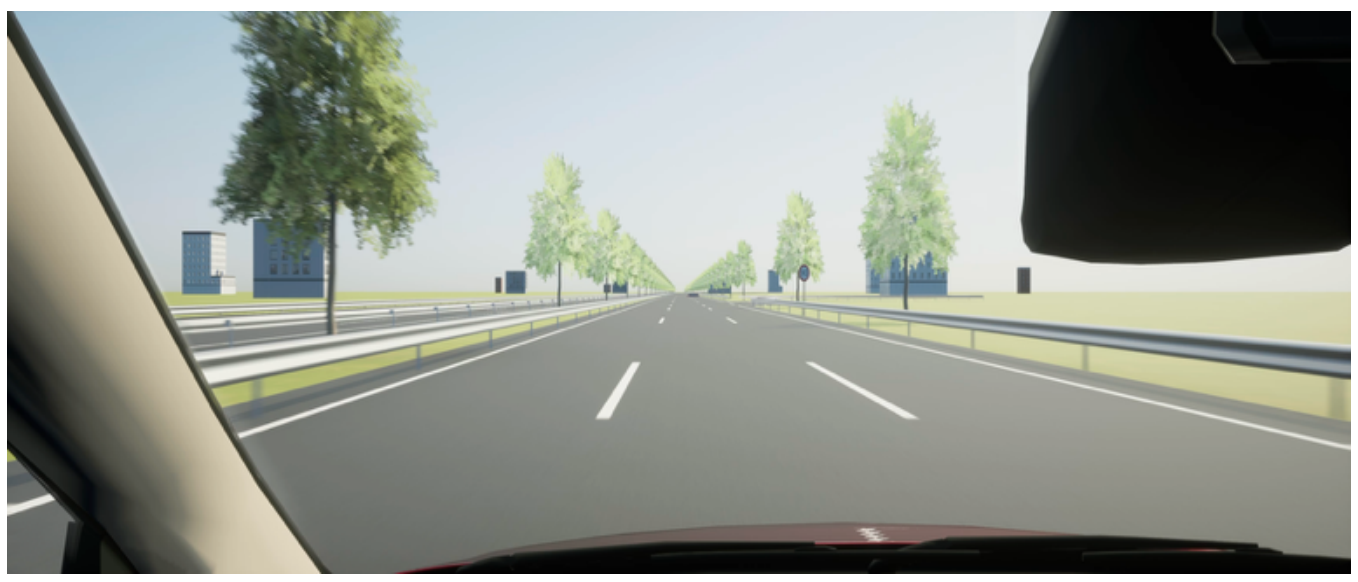

Figure 2. First-person perspective of the simulated highway driving scene used as the visual stimulus.

For each trial, two video clips were presented sequentially: one at the reference constant speed and one at a slightly different comparison speed. In the reference condition, the car speed was fixed at either 40 or 100 km/h. In the comparison clip, the speed differed by a certain $\Delta v$, which was adjusted adaptively from trial to trial. The order of presentation was counter-balanced, such that in half of the trials the reference-speed video was played first, and in the other half the comparison-speed video was presented first.

Three auditory conditions were tested in combination with the videos:

1. ICEV sound: a recording of an internal combustion engine vehicle, including engine noise, exhaust noise, and wind and tire noise corresponding to the vehicle speed.
2. EV sound: a recording of an electric motor sound, played together with the same wind and tire noise component.
3. Silence: no sound was provided.

For each stimulus, the visual and auditory cues were both speed dependent and corresponded to the visual and acoustic information produced by the tested vehicle type during constant-speed driving at the given speed.

### *Procedure*

Each participant was tested in all auditory conditions and at both base speeds (40 km/h and 100 km/h). The order of conditions was counterbalanced across participants to mitigate learning or order effects. For example, some participants experienced the "ICEV" sound condition first, while others started with "EV" or "Silence", in a Latin square rotation.

In each trial, the participant watched the two video segments (one reference and one comparison) in succession. After viewing both, they were asked to judge which of the two seemed to be at a higher speed. This was essentially a two-alternative forced choice (2AFC) task with respect to the event to the speed difference. Participants indicated their choice using a mouse click on a dialog box that appeared during the inter-stimulus interval.

An adaptive staircase method was employed to adjust the speed difference $\Delta v$ between the reference and comparison videos based on the participant's responses. We used a "2-down/1-up" staircase rule combined with Parameter Estimation by Sequential Testing (PEST) adaptive step-size adjustments [27]. Specifically, if the participant answered correctly (i.e., identified the faster video) in two consecutive trials, the next $\Delta v$ was decreased, making the task more difficult. If the participant answered incorrectly on a trial, $\Delta v$ was increased to make the difference easier to detect. The step size for adjusting $\Delta v$ followed PEST rules: the step size was halved whenever the direction of change reversed (a reversal indicates the point where responses switch from correct to incorrect or vice versa), and the step size was doubled if three consecutive changes occurred in the same direction without a reversal. Predefined minimum and maximum step sizes were set to prevent the difference from becoming too small or excessively large. To ensure that participants could clearly perceive the speed difference at the beginning of each block and to facilitate faster convergence of the adaptive procedure (thus reducing the total number of trials and minimizing participant fatigue), the initial comparison speed offset was set to ±10 km/h in the 40 km/h condition and ±32 km/h in the 100 km/h condition. The minimum step size for adjusting $\Delta v$ was set to 1 km/h for the 40 km/h base speed, and 2 km/h for the 100 km/h base speed, respectively. Each condition's staircase continued until one of the termination criteria was met: either 5 reversals occurred after the minimum step size was reached (indicating convergence), or 50 trials were completed, or at least 12 reversals in response trend occurred. To reduce fatigue, each block was followed by a mandatory 1-minute pause, after which participants could extend the break at their discretion before starting the next block.

Throughout the experiment, data from each trial were recorded. This included the base speed, the current $\Delta v$ and resulting comparison speed, the order of presentation (which video came first), the participant's choice of which seemed faster, and whether that choice was correct relative to the actual speeds. These data were later used to compute the JND thresholds and perform statistical analysis.

### *Data Analysis*

The just noticeable difference (JND) for each participant and auditory condition was derived directly from the reversal history of the adaptive staircase. For each staircase block, we retained the last six reversals (or all available reversals if fewer than six occurred) and recorded the stimulus levels immediately preceding them. To guard against the possibility that an observer experienced at least one extended lapse of attention, we trimmed the most extreme reversal: the absolute deviation of every reversal from the sample median was calculated, and any point whose deviation exceeded twice the mean of these deviations was discarded [28]. The JND was then taken as the arithmetic mean of the remaining reversals.

The staircase algorithm always reached a formal stopping rule. However, at the reference speed of 40 km/h, all 30 participants completed the experiment under the three auditory conditions, yielding a total of 90 staircase runs. Among these, 10 runs did not converge to a stable reversal level. Of these 10 non-convergent runs, nine occurred in the Silence condition and one occurred in the EV sound condition. Notably, these 10 unstable runs were distributed across 10 different participants. To ensure a balanced between-condition comparison at the same speed level, all runs at 40 km/h from these 10 participants were excluded from further analysis. By contrast, at 100 km/h every staircase run converged cleanly in all sound conditions and provided a usable JND. We interpret the low-speed instability as a consequence of the very weak optic-flow signal at 40 km/h: without engine or motor noise this visual cue is often insufficient for reliable speed discrimination, which explains the high failure rate in Silence. Nevertheless, the fact that a few staircases also failed for the same participants under ICEV or EV sound indicates that individual lapses of attention, compounded by the small absolute step sizes used at the low speed, can still destabilize the adaptive algorithm even when some auditory feedback is present.

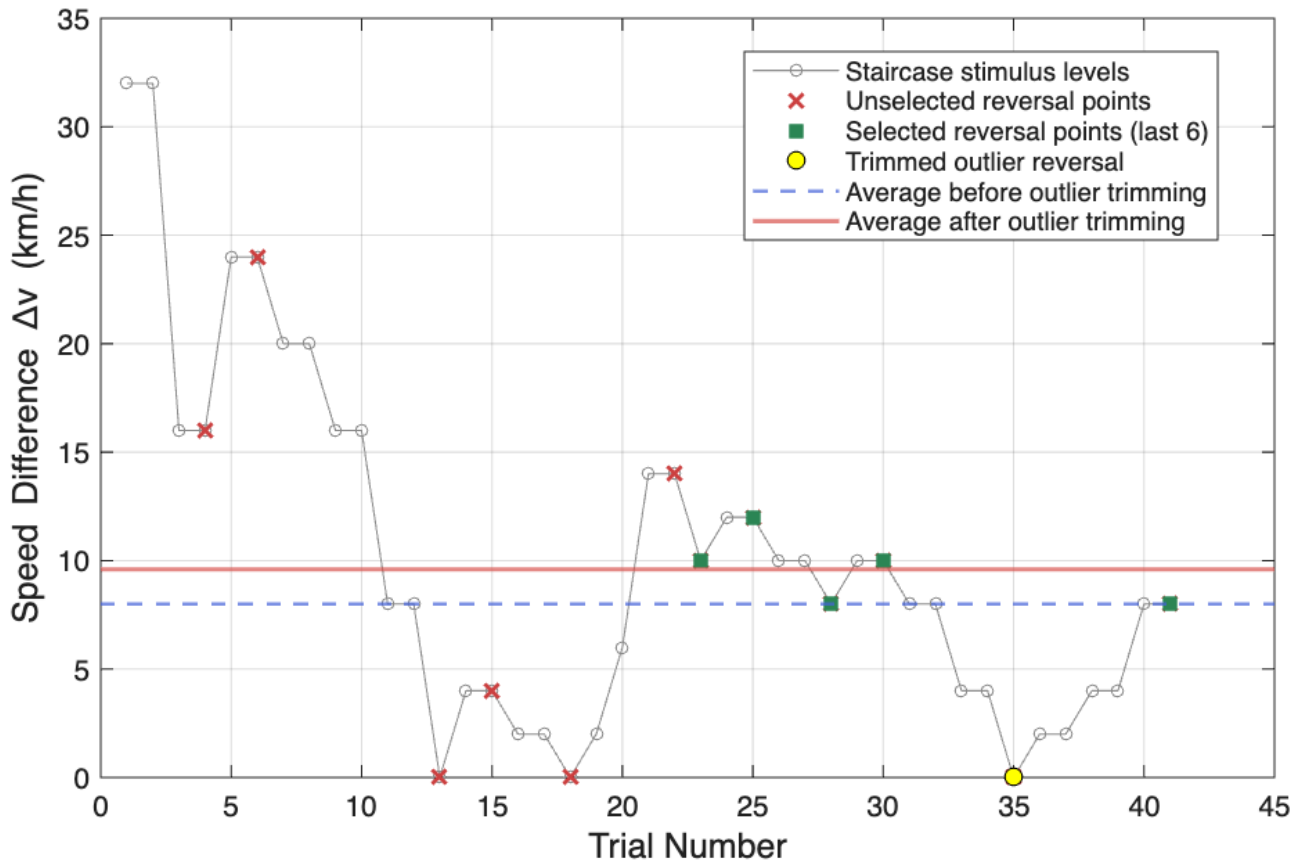

Figure 3. Complete staircase run from one participant in the 100 km/h condition. The plot shows the adaptive adjustment of the speed-difference stimulus (Δv) across trials and the reversals used to estimate the discrimination threshold. The just-noticeable difference (JND) is computed as the mean of the last several reversal levels, with one reversal removed when identified as an outlier using a median-deviation criterion. The dashed and solid horizontal lines represent the raw and outlier-pruned JND estimates, respectively.

For the condition-wise JND dataset, potential outliers were removed before any descriptive or inferential statistics were conducted. Within each auditory–speed condition, a participant's JND was flagged as an outlier if it lay more than 1.5 interquartile ranges (IQR) above the third quartile or below the first quartile. This trimming was applied only to the JND values and is conceptually distinct from the convergence-based exclusions described earlier. All summary statistics and hypothesis tests reported below are based on the resulting cleaned dataset.

To evaluate the effects of sound and speed conditions on JND thresholds, a within-subject statistical analysis was conducted. A two-way repeated-measures analysis of variance was conducted to assess the main effects of sound condition and speed level, as well as their interaction. The sound condition comprised ICEV, EV, and silence, and the speed levels were 40 km/h and 100 km/h.Post-hoc pairwise comparisons between conditions were performed using paired t-tests with Bonferroni-corrected significance thresholds. In addition, a non-parametric bootstrap resampling procedure (10,000 iterations) was applied to estimate 95% confidence intervals for condition-wise differences in JND and to verify the stability of t-test results under minimal distributional assumptions. A significance threshold of $\alpha = 0.05$ was used for all inferential tests.

## Results

Figure 4 displays the updated distribution of just-noticeable-difference (JND) values, with statistical outliers removed, for the three auditory conditions (ICEV, EV, Silence) at the two reference speeds (40 km/h and 100 km/h). Descriptive statistics are summarised in Table 1. At 40 km/h the mean JNDs were 1.24 km/h for ICEV, 1.19 km/h for EV, and 2.72 km/h for Silence. At 100 km/h the corresponding means rose to 1.93, 3.48, and 5.15 km/h, respectively, confirming that sensitivity to speed changes decreases as the baseline speed increases.

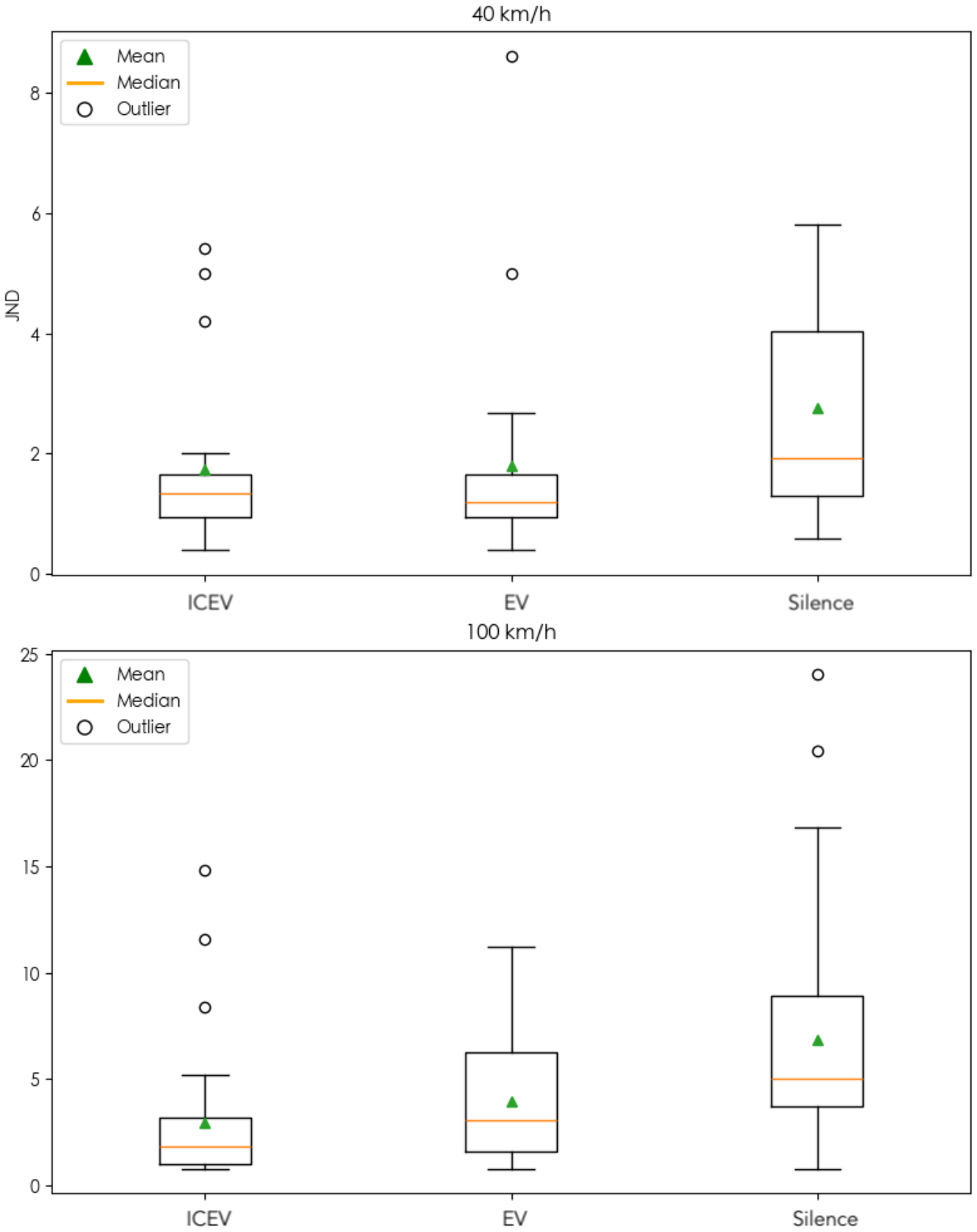

Figure 4. Distribution of Just Noticeable Difference (JND) across Auditory Conditions at Two Speed Levels (Outliers Excluded). Boxplots show medians (orange lines), means (green triangles), and outliers (open circles).

Table 2. Average of JND after outlier removal. (km/h)

| Condition | 40 km/h | 100 km/h |
|---|---|---|
| ICEV | 1.24 | 1.93 |
| EV | 1.19 | 3.48 |
| Silence | 2.72 | 5.15 |

A one-way repeated-measures ANOVA conducted separately at each speed level revealed a reliable main effect of sound condition at 40 km/h, $F(2,28) = 8.96$, Greenhouse–Geisser–corrected $p = 0.006$, $\eta^2 G = 0.31$, and likewise at 100 km/h, $F(2,50) = 10.07$, corrected $p = 0.0007$, $\eta^2 G = 0.24$, after adjusting for violations of sphericity in both cases. Bonferroni-adjusted pairwise tests showed that at 40 km/h Silence produced significantly larger JNDs than either ICEV or EV ($p < 0.01$ for both comparisons), while ICEV and EV did not differ ($p > 0.80$). At 100 km/h all three pairwise contrasts reached significance (ICEV < EV < Silence; ICEV vs. EV and ICEV vs. Silence, $p < 0.001$; EV vs. Silence, $p < 0.05$). Non-parametric bootstrap confidence intervals (10,000 resamples) confirmed the same pattern of effects.

Taken together, the analysis reinforces the conclusion that removing internal combustion engine noise degrades drivers' sensitivity to speed changes. Silence produced the largest JNDs at both reference speeds, and EV sound provided only partial improvement (matching ICEV at 40 km/h but lagging at 100 km/h).

## Discussion

Visual cues typically predominate in the perception of absolute speed; however, auditory inputs can considerably influence a driver's sensitivity to speed fluctuations. Our findings support this concept: the absence of engine or motor sounds resulted in notably poorer detection of speed changes, highlighting a significant cross-modal impact of auditory cues on speed discrimination.

### *Effects of Removing Auditory Feedback (Silence vs. Sound)*

One evident observation is that the elimination of auditory feedback significantly impairs speed discrimination abilities. In the "Silence" condition, characterized by the absence of engine or e-motor noise, participants required a considerably larger speed alteration to perceive any difference compared to scenarios where vehicle sounds were present. Quantitatively, the just-noticeable difference (JND) for speed, approximately doubled in the silence condition compared to conditions with sound at both measured speeds of 40 km/h and 100 km/h. Notably, at the slow speed of 40 km/h, numerous participants found it challenging to detect minor speed changes without sound and several could not establish a stable threshold at all during the "Silence" condition. This instability was due to the low optic flow at 40 km/h, where the visual scene undergoes minimal change with small speed variations, thus without any auditory cue, the task approached the boundaries of human perception. Consequently, in Silence at 40 km/h, the adaptive tracking procedure frequently yielded erratic outcomes, because of lack of clear visual as well as acoustic cues. Conversely, whenever engine or motor sounds were available, even at 40 km/h, participants showed significantly increased sensitivity and the estimation processes converged to a consistent JND. In summary, auditory feedback provided an essential sensory cue at low speeds, affirming that visual perception alone is insufficient for precise speed assessments at lower velocities.

At 100 km/h, visual cues to speed (optic flow, roadside passing rate) are stronger in an absolute sense, yet we still found that lack of sound impaired performance significantly. Every participant was able to complete the task at 100 km/h, but the mean JND in Silence was about 5.15 km/h, versus ~1.9–3.5 km/h with sound present (see next section for sound type differences). Thus, even at highway speed, drivers required over 5 km/h of change to detect a speed difference without sound, compared to much smaller changes when auditory cues were available. This aligns with prior studies showing that people tend to underestimate their speed when auditory feedback is reduced, leading them to require a larger deviation to feel a difference and often causing them to inadvertently drive faster.

### *Comparing ICEV Noise and EV Noise*

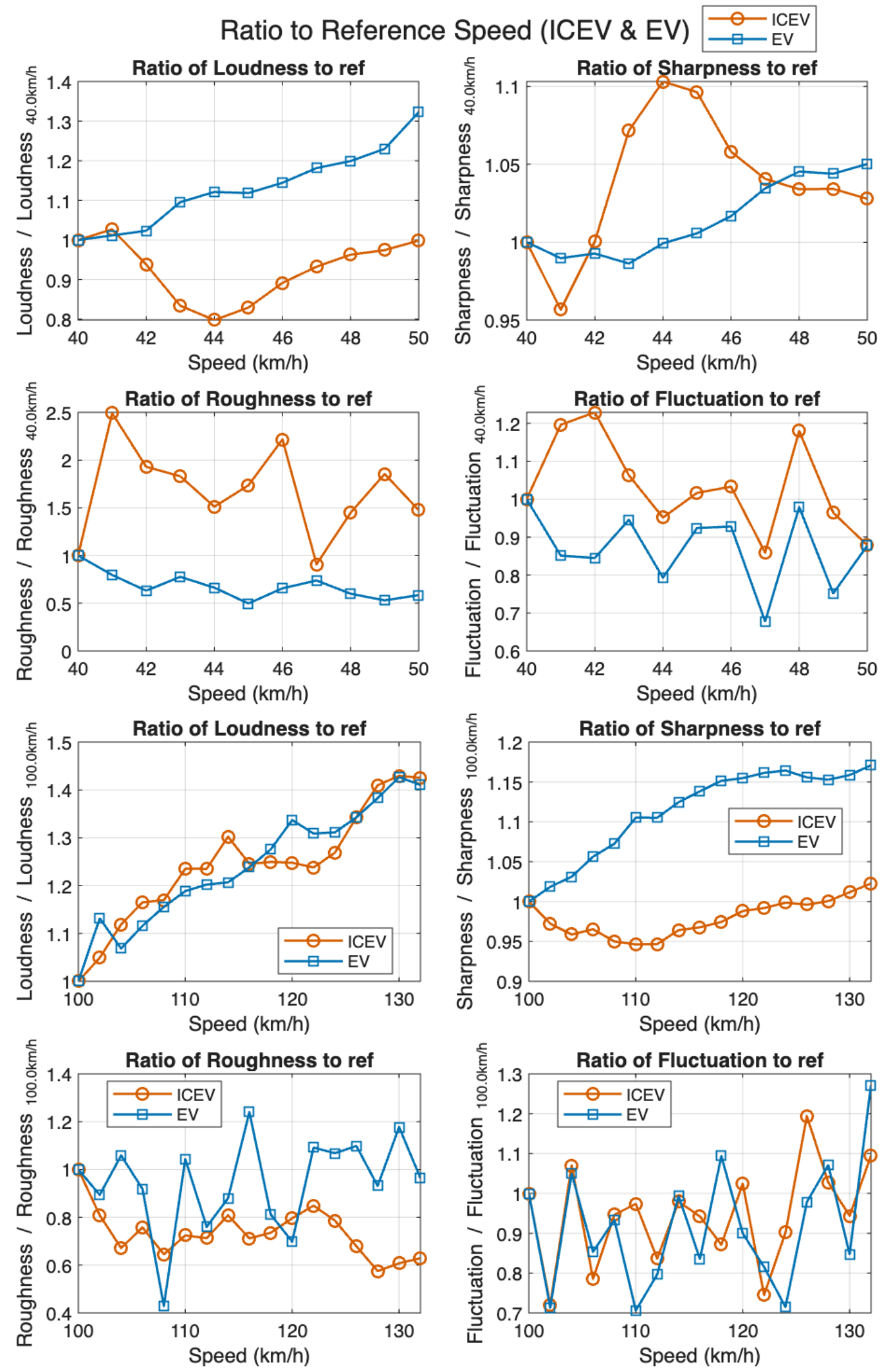

Figure 5. Ratios of psychoacoustic metrics relative to a reference speed (ICEV vs. EV)

Although any auditory cue is preferable to none for speed perception, differences in in-cabin sounds associated with different powertrain types do affect speed perception JNDs. At 40 km/h, JNDs did not differ between ICEV and EV, which is consistent with the 40 km/h normalized plots: For the sounds corresponding to the JND speed and the reference speed, the loudness ratios in both vehicles remain within the ±10% band and are therefore not considered perceptually detectable differences. In contrast, the ICEV exhibits pronounced variations in roughness, whereas the EV shows little to no such variation. When the overall sound level is close to the auditory threshold, these roughness variations may provide additional temporal information; however, at this speed they do not translate into a measurable difference in JND. At 100 km/h, the ICEV and EV differ in JND despite similar psychoacoustics metrics trends. Loudness ratios remain close to the ±10% detectability band for both vehicles, while roughness and fluctuation strength show no stable monotonic increase with speed. Sharpness increases more for the EV, but this wind/tyre "hiss" changes too weakly with speed to track threshold-level increments. Thus, the difference between EV and ICEV at highway speed cannot be well explained by these four psychoacoustic metrics alone.

Based on the tonality analysis using the tone to noise ratio shown in Figure 6, this difference can be explained directly. At 40 km/h, neither EV nor ICEV exhibit clear tonal components, indicating that their spectral structures are mainly dominated by broadband noise and that no stable or salient tonal cues are present. At 100 km/h, however, under the ICEV condition, the tone to noise ratio clearly identifies multiple prominent discrete tones concentrated around approximately 140 Hz, with a tonal level reaching about 6 dB. This indicates that, under high-speed conditions, the engine-order harmonics of the ICEV are perceptually stable and salient. In contrast, the EV signal at the same speed remains entirely dominated by broadband noise. This difference directly explains the divergence in human speed-perception performance between EV and ICEV sounds under high-speed driving conditions. Considering the relationship between engine rotational speed and engine-order frequencies, at a vehicle speed of 100 km/h, the engine speed is approximately 4300–4400 rpm. For a four-stroke inline four-cylinder engine, the second engine is typically the dominant order. Accordingly, the second-order frequency

$$f_{(2nd)} = 2 \times RPM / 60$$

is very close to the approximately 140 Hz component identified in the recording analysis, supporting the interpretation of this spectral peak as a second-order engine harmonic. Furthermore, using the measured engine speeds at 100 and 103 km/h (4350 and 4383 rpm). This second-order frequency $f_{100\,\mathrm{km/h}} \approx 146.1\,\mathrm{Hz}$ and $f_{103\,\mathrm{km/h}} \approx 145.0\,\mathrm{Hz}$, i.e., $\Delta f \approx 1.1\,\mathrm{Hz}$ or a fractional change $\Delta f / f \approx 0.76\%$. Classical frequency-discrimination limits around 150 Hz at moderate levels are on the order of 0.5 – 1.5%, so the speed-induced spectral shift over 3 km/h is at threshold or slightly above. The local mapping $df/dv \approx (f_{103} - f_{100})/3 \approx -0.37\,\mathrm{Hz}$ per (km/h) then predicts that the observed ICEV JND at 100 km/h ($\Delta v_{\mathrm{JND}} \approx 1.93\,\mathrm{km/h}$) corresponds to $\Delta f \approx 0.71\,Hz$, i.e., $\Delta f / f \approx 0.49\%$, a value squarely within many listeners' frequency-JND range. The agreement between the mapped frequency change (Δf) and classical frequency JNDs suggests that the ICEV speed JND at highway speed is consistent with a frequency-based cue. In other words, the harmonic components associated with the dominant engine order of an internal combustion engine are clearly audible at high driving speeds, and their frequencies shift continuously and predictably with vehicle speed, thereby providing listeners with a reliable speed-related auditory cue.

By contrast, under the same conditions, the EV lacks significant tonality, and its spectrum does not contain discrete components that can be stably tracked. As a result, speed changes are difficult to perceive through pitch- or order-related cues. This contrast indicates that, at high speeds, the advantage of ICEV in speed discrimination primarily arises from the engine-order tonal structure revealed by tone to noise ratio, rather than from simple differences in loudness. At low speeds, other auditory cues such as roughness or broadband noise variations may contribute to speed discrimination to some extent, but they are insufficient to compensate for the absence of tonal cues at high speeds.

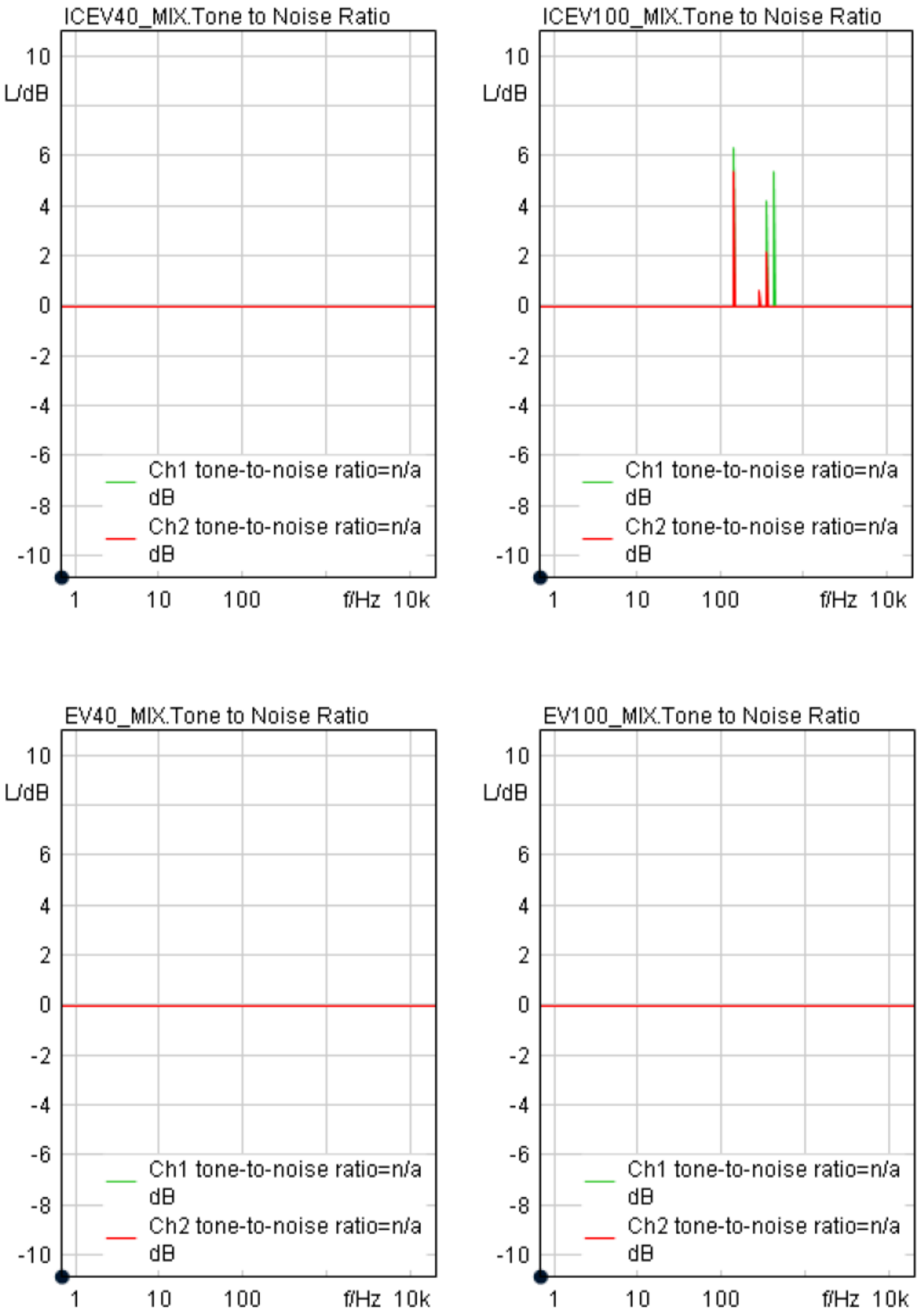

Figure 6. Comparison of Tone to Noise Ratio for EV and ICEV at 40 km/h and 100 km/h. Ch1 and Ch2 denote left- and right-ear recordings, respectively.

### *Limitations and Future Work*

The participant sample was relatively homogeneous in age, consisting primarily of engineering students aged between 20 and 25 years. In addition, participants were generally more familiar with ICEVs, while their limited experience with EVs may have introduced a potential bias. Future work should recruit a more diverse and representative participant sample in terms of age, background, and driving experience to improve the generalizability of the findings.

Future work could also extend the present approach by incorporating additional sensory modalities that are available in real vehicles, such as haptic feedback from the seat and steering wheel. These cues are strongly influenced by road excitation and vehicle dynamics and are therefore inherently speed dependent, making them potentially informative for speed perception. Moreover, investigating more dynamic driving scenarios, including richer visual environments and time-varying speed profiles (e.g., acceleration and deceleration phases), could help determine whether EV sound signatures primarily lead to speed underestimation or whether overestimation may also occur under certain conditions.

## Conclusion

This study investigated drivers' sensitivity to speed changes under different interior sound conditions using a psychophysical approach. The differences were quantified using just noticeable differences, and their relationships with acoustic cues were analyzed. Overall, the results confirm that removing auditory feedback degrades speed discrimination, but they further show that the diagnostic value of auditory cues depends on both vehicle speed and powertrain type.

Differences in speed perception between internal combustion engine vehicles and electric vehicles have been suggested in previous studies; the present results provide direct evidence supporting this view, with the disparity becoming more pronounced at higher speeds. At low speed (40 km/h), speed-discrimination performance did not differ between ICEV and EV conditions. At high speed (100 km/h), clear differences emerged between powertrain types. ICEV sounds yielded significantly lower JNDs than EV sounds, a mismatch that cannot be explained by loudness- or sharpness-related cues alone. Instead, tonality analysis revealed that ICEV interior noise contains stable, prominent engine-order harmonics whose frequencies shift systematically with vehicle speed. These tonal components provide a trackable spectral cue that supports fine-grained speed discrimination. By contrast, EV interior sounds lack salient tonality and remain dominated by broadband noise, making speed changes more difficult to perceive through pitch or order-based cues.

For electric vehicles, this suggests that simply increasing loudness is unlikely to improve speed perception. Instead, introducing stable tonal components that vary with speed, while maintaining acoustic comfort, may represent a more effective strategy for supporting drivers' speed perception and control.

The experiment and analysis code are openly available on GitHub at https://github.com/ZhenxianLi/Driving-Speed-JND (MIT License); the audio and video stimuli are permanently archived on Zenodo at https://doi.org/10.5281/zenodo.20617493 (CC BY 4.0).

## Contact Information

Zhenxian LI

zhenxian.li@insa-lyon.fr

INSA de Lyon Bâtiment St. Exupéry 25 bis av. Jean Capelle

69621 Villeurbanne cedex – France

## Acknowledgments

This work was supported by the European Commission through the Horizon Europe Programme (GAP-Noise project, with Grant Agreement No. 101073014)